\documentclass{iaus}
\usepackage{graphicx}
\usepackage{subfigure}

\title[Solar System--Debris Disk Connection] %% give here short title %%
{On the Solar System--Debris Disk Connection}

\author[Amaya Moro-Mart\'{\i}n]   %% give here short author list %%
{Amaya Moro-Mart\'{\i}n}

\affiliation{Department of Astrophysical Sciences, Princeton University, \\
Princeton, NJ 08544, USA\\ e-mail: {\tt amaya@astro.princeton.edu}}

\pubyear{2008}
\volume{249}  %% insert here IAU Symposium No.
\pagerange{1--10}
 \jname{Exoplanets: Detection, Formation and
Dynamics} \editors{Y.-S. Sun, S. Ferraz-Mello \& J.-L. Zhou, eds.}
\begin{document}

\maketitle

\begin{abstract}
This paper emphasizes the connection between solar and extra-solar debris disks:
how models and observations of the Solar System are helping us understand the debris 
disk phenomenon, and vice versa, how debris disks are helping us place our 
Solar System into context. 
\keywords{asteroids -- cirumstellar matter -- infrared: stars -- Kuiper Belt -- planetary systems -- Solar System.}
\end{abstract}

\firstsection % if your document starts with a section,
\section{Introduction}

Debris disks are disks of dust 10s--100s AU in size that surround main sequence stars of a wide range of 
stellar types (A to M) and ages (0.01--10 Gyr). In general, debris disks are not spatially 
resolved and are identified in the infrared from the dust thermal emission that results in an excess 
over the expected stellar values. Debris disks surveys carried out with {$\it Spitzer$} 
indicate that they contain a few lunar masses of dust and negligible quantities of gas, and that they 
are present around $>$33\% of A-type stars (Su  {\it et al.} 2006) and 10--15\% of solar-type FGK stars 
(Bryden  {\it et al.} 2006; Beichman  {\it et al.} 2006; Trilling  {\it et al.} 2008; Hillenbrand {\it et al.} 
in preparation; Carpenter  {\it et al.} in preparation).   However, these results are 
calibration limited because the disks can only be detected at a certain level above the 
stellar photosphere due to uncertainties in the stellar flux.
Figure 1 shows examples of some nearby spatially resolved debris disks. 

\begin{figure}
\begin{center}
 \includegraphics[angle=0,scale=0.62]{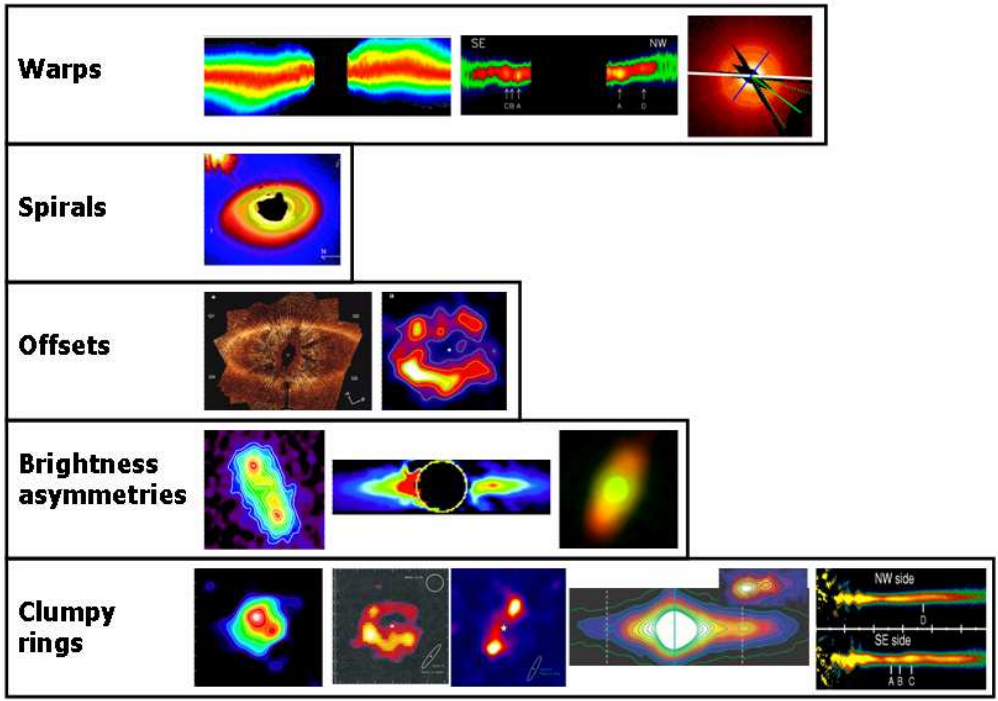}
\end{center}
\caption{\small Spatially resolved images of nearby debris disks showing a 
wide diversity 
of debris disk structure. From left to right the images correspond to: 
($\it{1st~row}$) 
$\beta$-Pic (0.2--1~$\mu$m; Heap {\it et al.}, 2000), 
AU-Mic (1.63 ~$\mu$m; Liu, 2004) and 
TW Hydra (0.2--1~$\mu$m; Roberge, Weinberger and Malumuth, 2005);
($\it{2nd~row}$) HD 141569 (0.46--0.72~$\mu$m; Clampin {\it et al.}, 2003); 
($\it{3rd~row}$) Fomalhaut (0.69--0.97~$\mu$m; Kalas {\it et al.}, 2005) and 
$\epsilon$-Eri (850~$\mu$m; Greaves {\it et al.}, 2005);
($\it{4th~row}$) HR4796 (18.2~$\mu$m; Wyatt {\it et al.}, 1999), 
HD 32297 (1.1~$\mu$m; Schneider, Silverstone and Hines, 2005) 
and Fomalhaut (24 and 70~$\mu$m; Stapelfeldt {\it et al.}, 2004); 
($\it{5th~row}$) Vega (850~$\mu$m; Holland {\it et al.}, 1998), 
$\epsilon$-Eri (850~$\mu$m; Greaves {\it et al.}, 1998), 
Fomalhaut (450~$\mu$m; Holland {\it et al.}, 2003),  
$\beta$-Pic (12.3~$\mu$m; Telesco {\it et al.}, 2005) 
and Au-Mic (0.46--0.72~$\mu$m; Krist {\it et al.}, 2005).
All images show emission from 10s to 100s of AU.}
\label{aba:fig1}
\end{figure}

The term ${\it debris}$ refers to the fact that the dust cannot be primordial, because the 
expected lifetime of the dust grains due to Poynting-Robertson drag ($ t_{PR} = 710 ({b \over \mu m}) 
({\rho \over g/cm^3})$ $({R \over AU})^2 ({L_\odot \over L_{star}}) {1 \over 1+albedo} ~yr, $
where $\it{R}$, $\it{b}$ and $\rho$ are the grain location, radius and density, respectively -- 
Burns, Lamy and Soter, 1979 and Backman and Paresce, 1993) and mutual grain collisions 
($  t_{col} = 1.26 \times 10^4 ({R \over AU})^{3/2} ({M_\odot \over M_*})^{1/2} 
({10^{-5} \over L_{dust}/L_*}) yr $ -- Backman and Paresce, 1993) 
is much shorter than the age of the star,  which means that the dust 
is likely being regenerated by planetesimals like the asteroids, Kuiper Belt objects (KBOs) 
and comets in our Solar System. 

Indeed, the Solar System is filled in with dust. The sources of dust are the asteroids and 
comets in the inner region and the KBOs and interstellar dust in the outer region. 
The dust produced in the inner region can be seen in scattered light with our naked eyes, either 
in the zodiacal light on in the coma of comets, and has extensively been observed in thermal 
emission by space-based observatories (${\it IRAS}$ and ${\it COBE}$). Evidence of the presence 
of dust originated in the Kuiper Belt (KB) comes from dust collision events detected by Pioneer 
10 and 11 beyond the orbit of Saturn (Landgraf {\it et al.}, 2002). Figure 2 shows the location 
of the planetesimals in the outer Solar System (left) and the expected spatial distribution 
of the dust generated in that region (right). 

\begin{figure}
\centering
\subfigure[Planetesimals] % caption for subfigure a
{
   \label{fig:sub:a}
    \includegraphics[width=5cm]{Outer_white.ps}
}
\hspace{1cm}
\subfigure[Dust] % caption for subfigure b
{
    \label{fig:sub:b}
    \includegraphics[width=5cm]{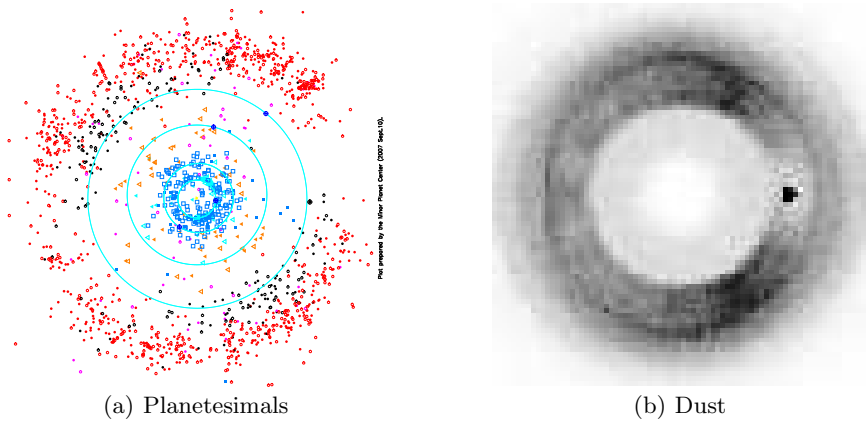}
}
\caption{(${\it Left}$) Distribution of planetesimals in the outer Solar System 
(courtesy of G. Williams at the Minor Planet Center). The outer circle is the orbit of 
Neptune. 
(${\it Right}$) Distribution of dust in the outer Solar System resulting from 
dynamical simulations of dust particles originated in the Kuiper Belt (from 
Moro-Mart\'{\i}n and Malhotra, 2002). The scale is the
same as in the previous panel, with the black dot representing the location of Neptune. 
The structure is the result of gravitational perturbations of the giant planets on 
the orbit of the dust particles (see Sec. 3).}
\label{fig:sub} % caption for the whole figure
\end{figure}

It is important to study the connection between the Solar System debris disks and 
the much brighter extra-solar debris disks because models and observations of the Solar System can help 
us understand the debris disk phenomenon, and vice versa, models and observations 
of extra-solar debris disks can help us place our Solar System into context. 

\section{Debris Disk Evolution}

\subsection{Steady collisional evolution}

It is thought that the Solar System was significantly more dusty in the past because 
both the Asteroid Belt (AB) and the Kuiper Belt (KB) were more densely populated. 
The system then became progressively less dusty as the planetesimal belts 
eroded away by mutual planetesimal collisions. Evidence of collisional evolution 
comes from the modeling and observation of the asteroid and KBO size distributions.
In the AB, Bottke {\it et al.} (2005) showed that the initial size distribution 
progressively changes from a power-law to the observed wavy distribution, with peaks at $D$$\sim$120 km
(leftover from the accretion process) and $D$$\sim$200 m (marking the transition at which 
the energy required to catastrophically destroy a particle is determined by self-gravity
rather than strength forces). 
In the KB, Bernstein {\it et al.} (2004) found that its current size distribution
shows a strong break to a shallower distribution at $D$$<$100 km (when particles become
more susceptible to collisional destruction). 

Models show that this collisional evolution likely resulted in the production of large quantities
of dust, as it can be seen in Figure 3 (from Kenyon and Bromley,
2005): in a planetesimal belt, Pluto-sized bodies ($D$$\sim$1000 km) excite the eccentricities of 
the smaller and more abundant 1--10 km sized planetesimals, triggering collisions and starting
a collisional cascade that produces dust and changes the planetesimal size distribution. Because 
the dust production rate is proportional to the number of collisions, and this is proportional to 
the square of the number of planetesimals, as the planetesimals erode and grind down to dust, the 
dust production rate decreases and the expected thermal emission from the dust slowly 
decays with time as $1/t$. This decay is punctuated by large spikes that are due to particularly 
large planetesimal collisions happening stochastically. Examples of stochastic events in the recent 
history of the Solar System are the fragmentation of the asteroids giving rise to the Hirayama and Veritas 
asteroid families (the latter happening 8.3 Myr ago and accounting for 25\% of the present zodiacal 
thermal emission; Dermott {\it et al.}, 2002) and the dust bands observed by {\it IRAS} 
(Sykes and Greenberg, 1986).

\begin{figure}
\begin{center}
 \includegraphics[angle=0,scale=0.2]{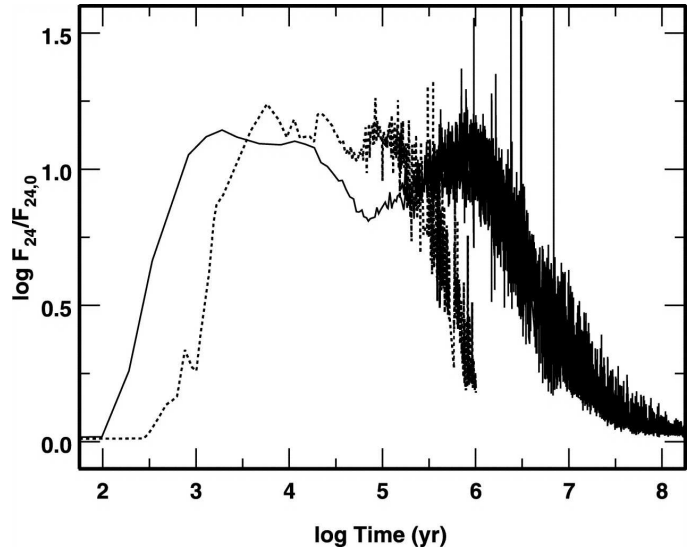}
\end{center}
\caption{\small Evolution with time of the 24 $\mu$m dust thermal emission expected from 
the collisional evolution of two planetesimal belts extending from 0.68--1.32 AU (dashed line) 
and 0.4--2 AU (solid line) around a solar type star (Kenyon and Bromley, 2005).
}
\label{aba:fig3}
\end{figure}

Recent surveys carried out by {\it Spitzer/MIPS} have enabled the detection of debris disks around 
hundreds of A-type and solar-type stars with a wide range of ages, showing that the dust emission 
follows a $1/t$ decay and there is a large variability likely due to individual collisions 
(see Figure 4), in broad agreement with the results from collisional cascade models (Su {\it et al.}, 2006; 
Siegler {\it et al.}, 2007) . Because solar and extra-solar planetary systems 
seem to follow similar evolutions, the imaging of debris disks at different 
evolutionary stages could be equivalent to a Solar System ``time machine''.

\begin{figure}
\begin{center}
 \includegraphics[angle=90,scale=0.45]{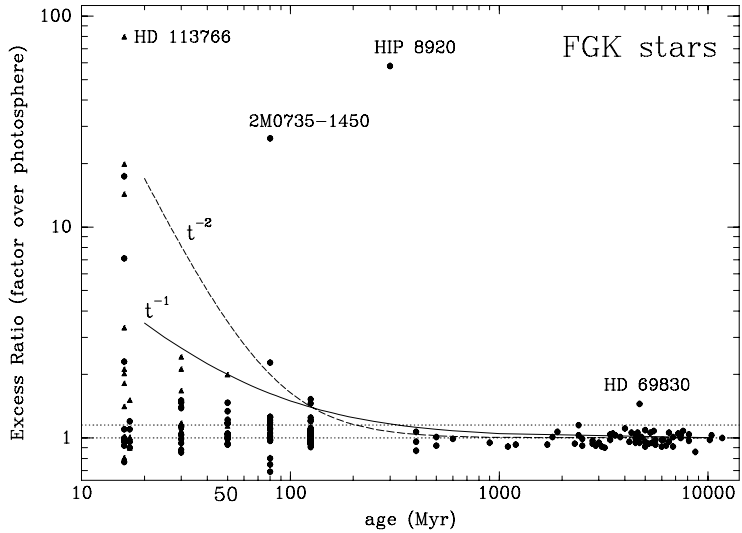}
\end{center}
\caption{\small Ratio of the dust emission to the expected stellar emission at 
24~$\mu$m for a survey of solar-type (FGK) stars. 
The stars aligned vertically belong to clusters or associations, 
therefore sharing the same age. The main features are the $1/t$ decay and the large variability
found for a given stellar age. A few
particularly massive debris disks are labeled. Figure from Siegler {\it et al.} (2007).}
\label{aba:fig4}
\end{figure}

\subsection{Stochastic non-collisional evolution}

As discussed above, there is observational and theoretical evidence that collisional evolution 
played a role in the evolution of solar and extra-solar debris disks. However, there is also 
evidence that additional non-collisional processes, likely related to the dynamical depletion 
of planetesimals that can result from gravitational interactions with massive planets, have also
played a major role is disk evolution. 

In the Solar System, evidence comes from the 
Late Heavy Bombardment (LHB, or Lunar Cataclysm), a period of time in the Solar System 
past during which a large number of impact craters in the Moon and the terrestrial planets were 
created (with an impact rate at Earth of $\sim$20000$\times$ the current value). 
This event, dated 
from lunar samples of impact melt rocks, happened during a very narrow interval of time -- 
3.8 to 4.1 Gyr ago ($\sim$600 Myr after the formation of the terrestrial planets). 
Thereafter, the impact rate decreased exponentially with a time constant 
ranging from 10--100 Myr (Chyba, 1990). Strom {\it et al.} (2005) compared the impact 
cratering record and inferred crater size distribution on the Moon, Mars, Venus and Mercury, 
to the size distribution of different asteroidal populations, showing that the LHB lasted 
$\sim$20--200 Myr, the source of the impactors was the main AB, and 
the mechanism was size independent. The most likely scenario 
is that the orbital migration of the giant planets caused a resonance sweeping of the 
AB and as a result many of the asteroidal orbits became unstable, causing a large scale 
ejection of bodies into planet-crossing orbits (explaining the observed cratering record), and
an increased rate of asteroidal collisions that would have been accompanied by the production of
large quantities of dust. Under this scenario, the LHB was a single event in the history of 
the Solar System (Strom {\it et al.}, 2005). 

A handful of extra-solar debris disks observed with {\it Spitzer} also show evidence of 
non-collisional evolution (e.g. BD+20307, HD 72905, eta-Corvi and HD 69830; 
Wyatt {\it et al.}, 2007). A particularly interesting case is that of 
HD 69830, a system that harbors three Neptune-like planets inside 2AU, shows a strong 24 $\mu$m dust 
emission indicative of large quantities of warm grains, no 70 $\mu$m dust emission indicative 
of a lack of cold dust, and a dust emission spectra with strong features remarkably similar to the 
spectra of comet Hale Bopp (Figure 5 -- Beichman {\it et al.}, 2005a). 
Wyatt {\it et al.} (2007) showed that its 24 $\mu$m emission, seen as an
outlier in Figure 4, implies a very high dust production rate that could not possibly have been  
sustained for the entire lifetime of the star and must therefore be a transient event rather 
than the results of steady collisional evolution. Its transient nature would account for the 
fact that HD 69830 shows strong silicate emission features indicative of the presence of large 
quantities of small grains. These grains have a very short lifetime and therefore must have been 
produced in a recent event. Wyatt {\it et al.}, 2007 suggested that we may be witnessing a LHB-type 
of event in which icy planetesimals originally located outside the orbit of the planets are scattered into the 
inner system producing the observed dust. 

\begin{figure}
\begin{center}
 \includegraphics[angle=0,scale=0.2]{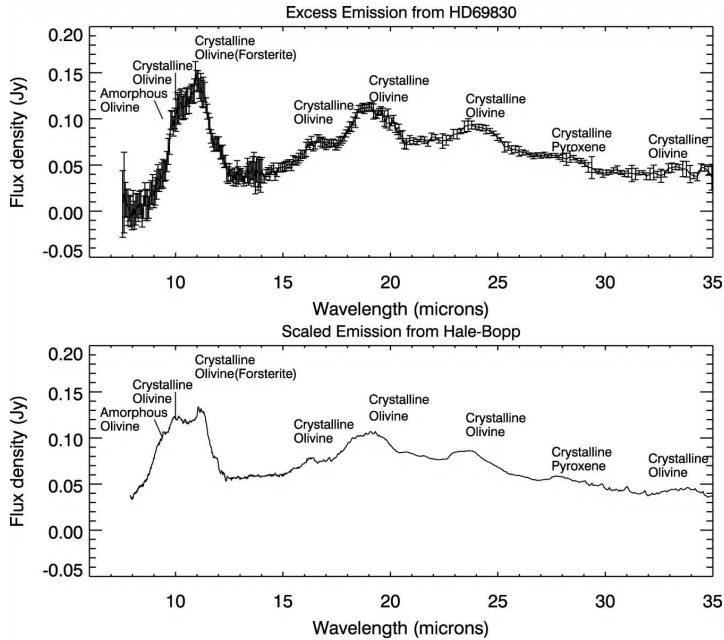}
\end{center}
\caption{\small Spectrum of the dust emission around HD 69830 ({\it top}) compared to the
spectrum of the comet Hale-Bopp normalized to a blackbody temperature of 400 K ({\it bottom}). Figure from 
Beichman {\it et al.} (2005a).}
\label{aba:fig5}
\end{figure}

The models and observations described above for both solar and extra-solar systems 
indicate that in a planetesimal swarm there is collisional evolution that produces dust, 
triggered by the largest (Pluto-sized) planetesimals in the swarm, and on top of that, 
depending on the planetary configuration, there may be drastic dynamical events that 
produce very significant depletion of planetesimals and an increased rate of planetesimal 
collisions and dust production. The next Section discusses how the presence of 
planets not only can affect the production of debris dust, but can also sculpt the debris 
disk by creating a rich diversity of spatial structure. 

\section{Debris Disk Structure}

Even though the great majority of debris disks observations are spatially 
unresolved, their structure can be studied in some detail through 
the spectral energy distribution (SED) of the disk because different 
wavelengths in the SED trace different distances to the star, so that 
an SED with sufficiently high spectral resolution can be used to constrain roughly the radial 
distribution of dust. Recent {\it Spitzer} debris disks surveys
suggest that debris disks commonly show evidence of the presence of inner cavities, 
as most systems show 70 $\mu$m dust emission (from cold dust), but no emission at $\lambda$$\leq$ 
24 $\mu$m (i.e., no warm dust;  see e.g. Meyer {\it et al.}, 2004; Beichman {\it et al.}, 2005b; 
Bryden {\it et al.}, 2006;  Kim {\it et al.}, 2005; Moro-Mart\'{\i}n, Wolf and Malhotra, 2005; 
Moro-Mart\'{\i}n {\it et al.}, 2007a;  Hillenbrand {\it et al.}, in preparation).  
High resolution spatially resolved observations have 
been obtained for a handful of nearby debris disks and indeed these images show the presence of 
inner cavities together with more complex morphology, like warps, 
spirals, offsets, brightness asymmetries and clumpy rings (see Figure 1). 

Dynamical simulations of the orbits of dust particles and their parent planetesimal in systems 
where massive planets are present suggest that this complex morphology could be the result from 
gravitational perturbations by planets 
(e.g. Roques {\it et al.}, 1994; Mouillet {\it et al.}, 1997; Wyatt {\it et al.}, 1999; 
Wyatt, 2005, 2006; Liou and Zook, 1999; Moro-Mart\'{\i}n and Malhotra, 
2002, 2003, 2005; Moro-Mart\'{\i}n, Wolf and Malhotra, 2005; Kuchner and Holman, 2003; 
see Moro-Mart\'{\i}n {\it et al.}, 2007b for a review). The basic mechanisms by which the
planets can affect the debris disks structure are the following:
\begin{itemize}
\item {\it Ejection by gravitational scattering:} This process can affect dust particles as 
they spiral inward under P-R drag, and dust-producing planetesimals, in the case when
the planet migrates outwards, resulting in a depletion of dust inside the orbit of the planet 
(an inner cavity). Dynamical simulations show that this process can be very efficient, 
ejecting $>$90\% of the particles in the case of a 3--10 M$_{Jup}$ planet
located between 1--30 AU around a solar-type star. 

\item {\it Trapping in mean motion resonances (MMR) with the planet:} In a system where the 
dust producing planetesimals are located outside the orbit of the planet, as the dust particle
drift inward due to P-R drag they can get trapped in MMRs with the planet. 
The MMRs are located where the orbital period of the planet is $\it{(p+q) / p}$ times 
that of the particle, where $\it{p}$ and $\it{q}$ are integers, $\it{p}$$>$0 and 
$\it{p+q}$$\geqslant$1. At these locations the particle receives energy from the perturbing 
planet that can balance the energy loss due to P-R drag, halting the inward motion of the 
particle and giving rise to planetary resonant rings. Due to the geometry of the resonance, 
the spatial distribution of material in resonance is asymmetric with respect to the planet, 
and this can explain the clumpy structure observed in some disks (Figure 1). 
An example of MMR trapping of KB in the Solar System can be seen in Figure 2, where
the ring-like structure, the asymmetric clumps along the orbit of Neptune, and the clearing 
of dust at Neptune's location are all due to 
the trapping of particles in MMRs with the planet, while the dust depleted region inside 10 AU 
is due to gravitational scattering by Saturn and Jupiter. MMRs can also affect the location 
of the planetesimals and the dust when the planets migrate outward. 

\item {\it Effects of secular perturbations:} These are the long-term average of the perturbing 
forces and act on timescales $>$0.1 Myr (see review by Wyatt {\it et al.}, 1999).
If the planet and the planetesimal disk are not coplanar, the secular perturbations tend to align 
the orbits and in the process they will create a warp in the disk. 
If the planet is in an eccentric orbit,  the secular perturbations will force an eccentricity on the 
dust particles, creating an offset in the disk center with respect to the star that can 
result in a brightness asymmetry. Other effects of secular perturbations are spirals and inner gaps. 
\end{itemize}

Finally, it is important to point out that because the debris disk structure is sensitive to the 
presence of planets located far from the star, the study of the structure could be used as a 
potential planet detection method that would be complementary to the well-established radial 
velocity and transit techniques (sensitive to close-in planets). 

\section{Concluding Remarks}

Large surveys of debris disks over a wide range of evolutionary states, enabled by 
high sensitivity spaced-based IR telescopes like {$\it Spitzer$}, are starting to 
provide a "movie" of how planetary systems evolve with time. In this regard, debris disks help us
place our Solar system into a broader context and vice versa, the study of the Solar System, 
in particular its dynamical history and the characterization of its small body population, 
sheds light on the physical processes giving rise to the debris disk phenomenon. Debris disks 
surveys, together with the results from planet searches, can help us understand the 
frequency of planetesimal and planet formation and the diversity of planetary systems, 
which ultimately addresses one of the most fundamental questions: 
is the Solar System common or rare?

\section*{Acknowledgments}
A.M.M. warmly thanks the organizers of the IAU Symposium No. 249 for their kind invitation 
and generous finantial support. 
A.M.M. is under contract with the Jet Propulsion Laboratory (JPL) funded by NASA through
the Michelson Fellowship Program. JPL is managed for NASA by the California Institute of
Technology. A.M.M. is also supported by the Lyman Spitzer Fellowship at Princeton University.

\end{document}